\documentclass[final,3p,times,twocolumn]{elsarticle}
\usepackage{blindtext}
\usepackage{enumerate}
\usepackage{commath}
\usepackage[italian,british]{babel}
\usepackage{amsmath}
\usepackage{textcomp}
\usepackage{upgreek}
\usepackage{braket}
\usepackage[T1]{fontenc} % It's an output font encoding, supporting most of northern european characters.
\usepackage[utf8]{inputenc} % Allows to input special characters from keyboard.
\usepackage{pdfpages}
\usepackage[version=4]{mhchem}
\usepackage{gensymb} % For degree symbol °
\usepackage{graphicx}
\usepackage{wrapfig} % Package used to wrap figures into the surrounding text
\usepackage{caption} % Manage captions in figures
\usepackage[font=small,textfont=it]{caption} % Figure font size
\usepackage{subcaption} %It is used to introduce multiple figures in a single caption.
\usepackage{float} % Allows further image positioning. Ex: [H]
\usepackage{booktabs}
\usepackage{fancyhdr}
\usepackage{epstopdf} % This package allows to con­vert eps images to pdf for use with pdfla­tex
\usepackage{soul} % For hyphenation, underlining, highlighting and more.
\usepackage{textcomp}
\usepackage{geometry}
\usepackage{enumitem}
\usepackage{amssymb}
\journal{NIM B}

\begin{document}

\begin{frontmatter}




\title{Material development towards a solid 100 kW electron-gamma converter for TRIUMF-ARIEL}


\author[TRIUMFaddress,UBC-CHEMISTRYaddress]{L. Egoriti}
\author[TRIUMFaddress,UVIC-PHYSICSaddress]{M. Cervantes}
\author[TRIUMFaddress]{T. Day Goodacre}
\author[TRIUMFaddress,UVIC-PHYSICSaddress]{A. Gottberg\corref{correspondingauthor}}
\cortext[correspondingauthor]{Corresponding author}
\ead{gottberg@triumf.ca}

\address[TRIUMFaddress]{TRIUMF, 4004 Wesbrook Mall, Vancouver, BC, V6T 2A3, Canada}
\address[UBC-CHEMISTRYaddress]{University of British Columbia, Department of Chemistry,2036 Main Mall Vancouver, BC, V6T 1Z1, Canada}
\address[UVIC-PHYSICSaddress]{University of Victoria, Department of Physics and Astronomy, Victoria, BC, V8W 2Y2, Canada}

\begin{abstract}
A series of irradiation tests have been performed at TRIUMF to investigate different material pairings to act as high-power electron-to-gamma converter for the ARIEL Electron Target East (AETE). The bulk of the converter body will be made out of an aluminum alloy with a sub-millimeter high-Z metal layer bonded to the surface facing the incoming electron beam. This contribution presents the approach chosen to select the optimal material for the high-Z layer, describes the tests performed and shows result which led to the successful selection of a specific tantalum-aluminum pairing as the future ARIEL converter material.
\end{abstract}

\begin{keyword}
ARIEL \sep electron-to-gamma converter \sep ISOL-target \sep radioisotope production
\end{keyword}

\end{frontmatter}

\section{Introduction}
\label{sec:Introduction}
The new ARIEL facility is being built to achieve a threefold improvement of Radioactive Ion Beam (RIB) hours available at the TRIUMF laboratory. The new ARIEL Proton Target West (APTW) will receive 500~MeV protons up to 50~kW from the existing cyclotron, while the new ARIEL electron linac will provide 100~kW, 30 MeV electrons to the ARIEL Electron Target East (AETE). This second target station not only is the first of its kind at a high-power ($>$500 W) ISOL facility but also exploits an independent driver beam, complementing the existing TRIUMF RIB production capabilities while increasing the reliability of RIB delivery \cite{ISACandARIEL}.
%A new 10 mA - 35 MeV electron Linac will provide a new driver beam for radioisotope production at the new Isotope Separation On-Line (ISOL) facility ARIEL which will improve by 200\% the production of Radioactive Ion Beams (RIBs) produced at TRIUMF by the existing ISAC facility .
The electron beam will be fully stopped in a converter body where it produces a shower of gamma rays with the intensity peak in the forward direction. A RIB production target is placed downstream in the photon path so that photonuclear reactions are induced in either actinide (for photofission) or lighter (typically for ($\gamma$,p) reactions) target materials. Their reaction products can be extracted from the target, ionized in an ion source, mass separated by two dipole magnets with combined design resolution 1:20,000 \cite{Maloney2016} and delivered to a variety of existing and future experiments in the ISAC-I and ISAC-II experimental halls. 

The production of radioisotopes using an electron beam has first been explored at CERN using the LPI injector LINAC \cite{Essabaa2003} and the promising results led to the construction of the ALTO facility which makes use of a 50 MeV, 500 W electron beam to induce the photofission reactions in an ISOLDE-type ISOL target \cite{Franchoo2015}. The ALTO targets are able to take the full 500 W electron beam power directly on target gaining a geometrical isotope production improvement with respect to the configuration where a tungsten converter was used to produce photons \cite{Essabaa2003}. At TRIUMF-ARIEL, the target itself can not sustain an electron beam power of 100 kW. Instead, the development of a converter body is necessary to fully stop the electrons and allow only the bremmstrahlung gamma rays to interact with the downstream ISOL target. A magnetic rastering system will scan the electron beam over the converter surface with a pre-defined pattern to maximize the electron beam current that the converter/target combination can accept by avoiding hot spots and protecting sensitive parts of the assembly.\\There are a number of factors relevant to the determination of the optimal converter material including: the electron-to-gamma conversion efficiency, the primary electron power deposition and dissipation in the converter and target bodies, the structural requirements for operation and the practicalities for manufacturing. Based on this, a composite material approach has been chosen which involves the deposition/bonding of a high-Z material layer onto a low-Z body such as aluminum. The former is supposed to maximize the production of gamma ray photons, while the supporting aluminum body would provide the necessary water-cooling and absorb the residual electrons as the optimized high-Z layer stops only about 50\% of the primary beam energy. Different converter designs such as a bare tantalum plate/foil were conceptually explored, but their combination of cooling performance and electron-to-gamma conversion were less efficient than the composite design. A more detailed description and justification of the TRIUMF-ARIEL converter design can be found in \cite{Cade2018}.

\section{Motivation for the Study} 

The most vulnerable aspect of the converter design is the contact interface between the aluminum and the high-Z material due to different thermal expansion between the two materials and the potential for chemical interactions at their interface. A delamination or disintegration of the high-Z layer would cause a chain reaction of overheating of the aluminum body, the target and finally the entire assembly. Among  metals with high atomic number, tantalum (Z=73) and gold (Z=79) were selected for their comparatively ductile nature which could comply with stresses at the interface with aluminum. On the one hand, the linear expansion coefficient of aluminum ($2.3 \cdot 10^{-5}$ m/m$\cdot$K) is closer to gold ($1.4 \cdot 10^{-5}$ m/m$\cdot$K) than tantalum ($6.5 \cdot 10^{-6}$ m/m$\cdot$K) and tungsten ($4.5 \cdot 10^{-6}$ m/m$\cdot$K) and therefore lower stresses and better mechanical performance were expected from this metal pairing. On the other hand, the formation of intermetallic phases between both Ta-Al \cite{Du1996} and Au-Al \cite{Okamoto2005} metals are known from literature but their formation dynamics and effects on the performance of the metal pairing for this specific application was still to be assessed.
The interplay of chemical and mechanical effects, the intrinsic presence of steep thermal gradients ($\approx$ 100 K/mm) and the criticality of the converter integrity for the success of the whole AETE target station called for the set up of a test station to investigate and ultimately select the optimal material pairing candidate for the converter materials. 

\section{Electron Irradiation Capabilities}
\begin{figure*}
	\centering
	\includegraphics[width=\linewidth]{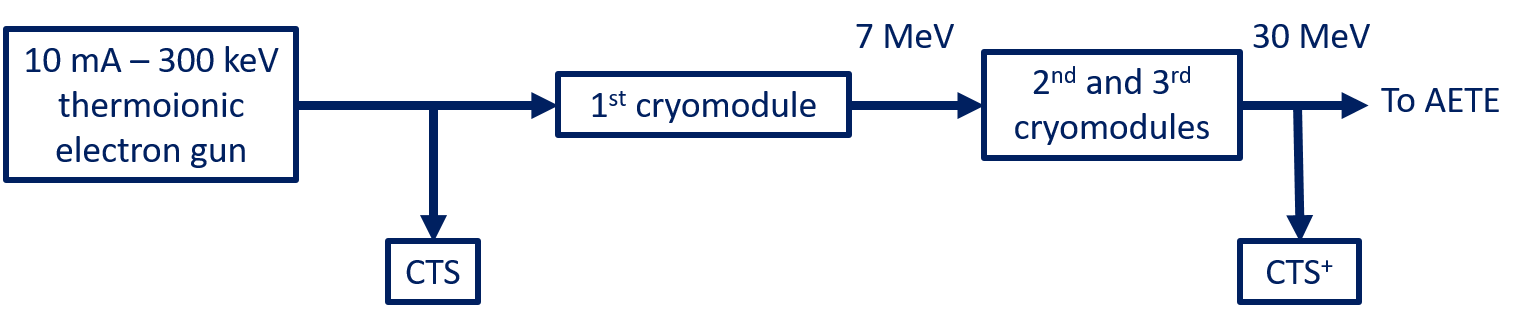}
	\caption{Irradiation capabilities of the ARIEL electron linac}
	\label{fig:ARIELlinac}
\end{figure*}

\begin{figure}
	\centering
	\includegraphics[width=0.5\linewidth]{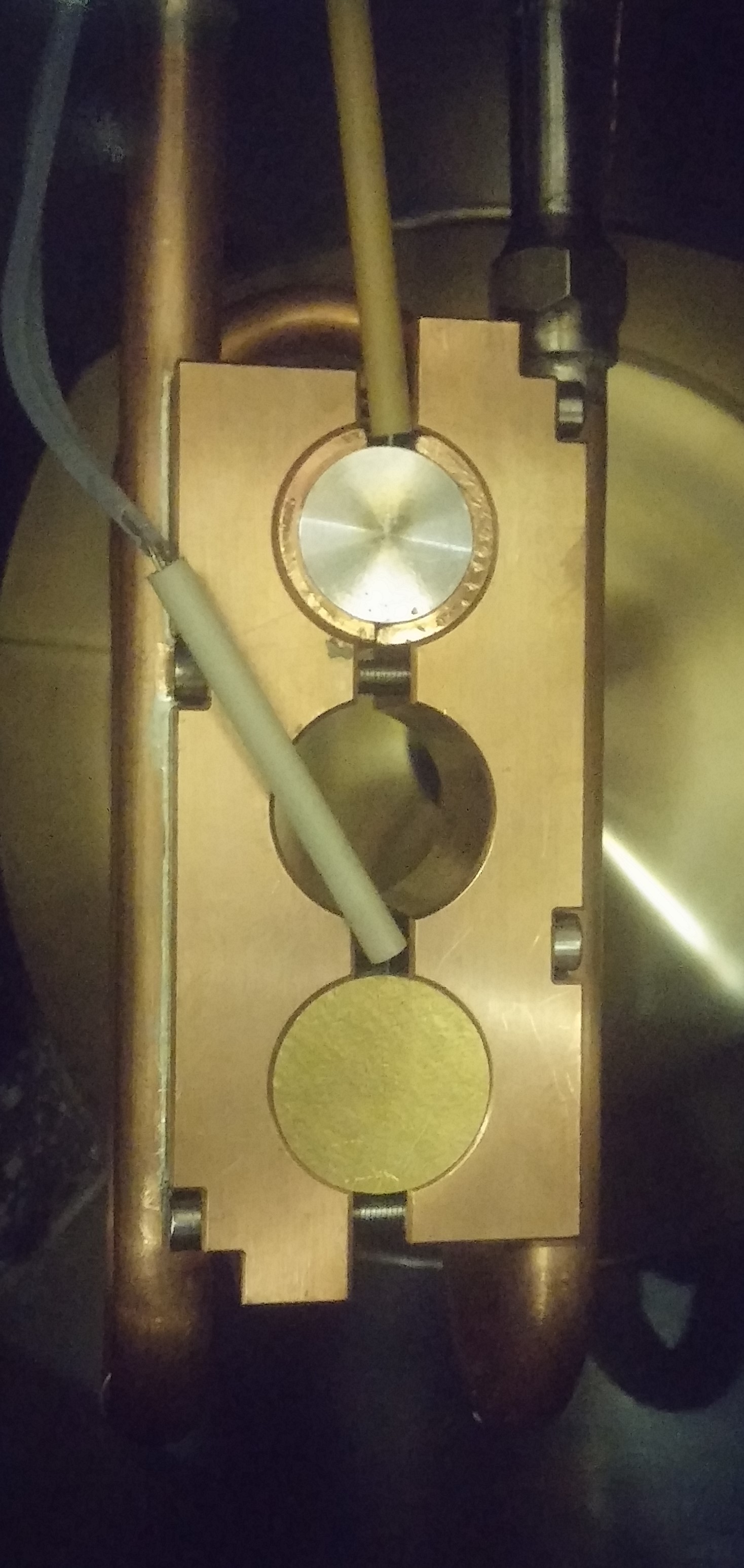}
	\caption{Water-cooled copper sample holder for high power irradiation studies equipped with a Au-Al sample and a dummy aluminum sample, with thermocouples installed to read temperature at the center of the samples.}
	\label{fig:CTS_SampeHolder}
\end{figure}

The 300 keV section of the ARIEL electron linac \cite{Koscielniak2017} highlighted in Fig. \ref{fig:ARIELlinac} is adapted to host the Converter Test Stand (CTS): a six-way vacuum cross equipped with remotely controlled sample insertion mechanisms and beam diagnostics. A water-cooled copper sample holder shown in Fig. \ref{fig:CTS_SampeHolder} is utilized to host up to three samples at a time and is equipped with a current readback to monitor the electron beam current delivered to a sample. 
Moreover, the cylindrical samples to be tested were precision drilled to host a thermocouple a fraction of a millimeter away from the interface between aluminum and the high-Z metal. Fig. \ref{fig:CTS_SampeHolder} shows the assembly hosting a gold-aluminum sample and a reference aluminum sample.
The electron beam power deposited in the actual AETE converter is approximately half of the incoming 100~kW driver beam ($\approx$~50~kW), which corresponds to a power density of $\approx$~10~W/mm$^2$ on the converter face of total area 50~cm$^2$. In order to match this power density, the 300 keV electron beam is impinged on the CTS sample at 1.6~mA current with a beam spot of 6~mm diameter (95~\% of beam). Future tests exploiting electron beam currents between 25~nA and 10~mA could be envisioned, where beam power up to~3 kW can be reached. Furthermore, a second lateral irradiation station (CTS$^+$) with higher beam energy up to 30~MeV will be installed as an upgrade of the existing high energy beam diagnostic box (see Fig. \ref{fig:ARIELlinac}).

\section{Experimental Technique}
\begin{figure}
	\centering
	\includegraphics[width=\linewidth]{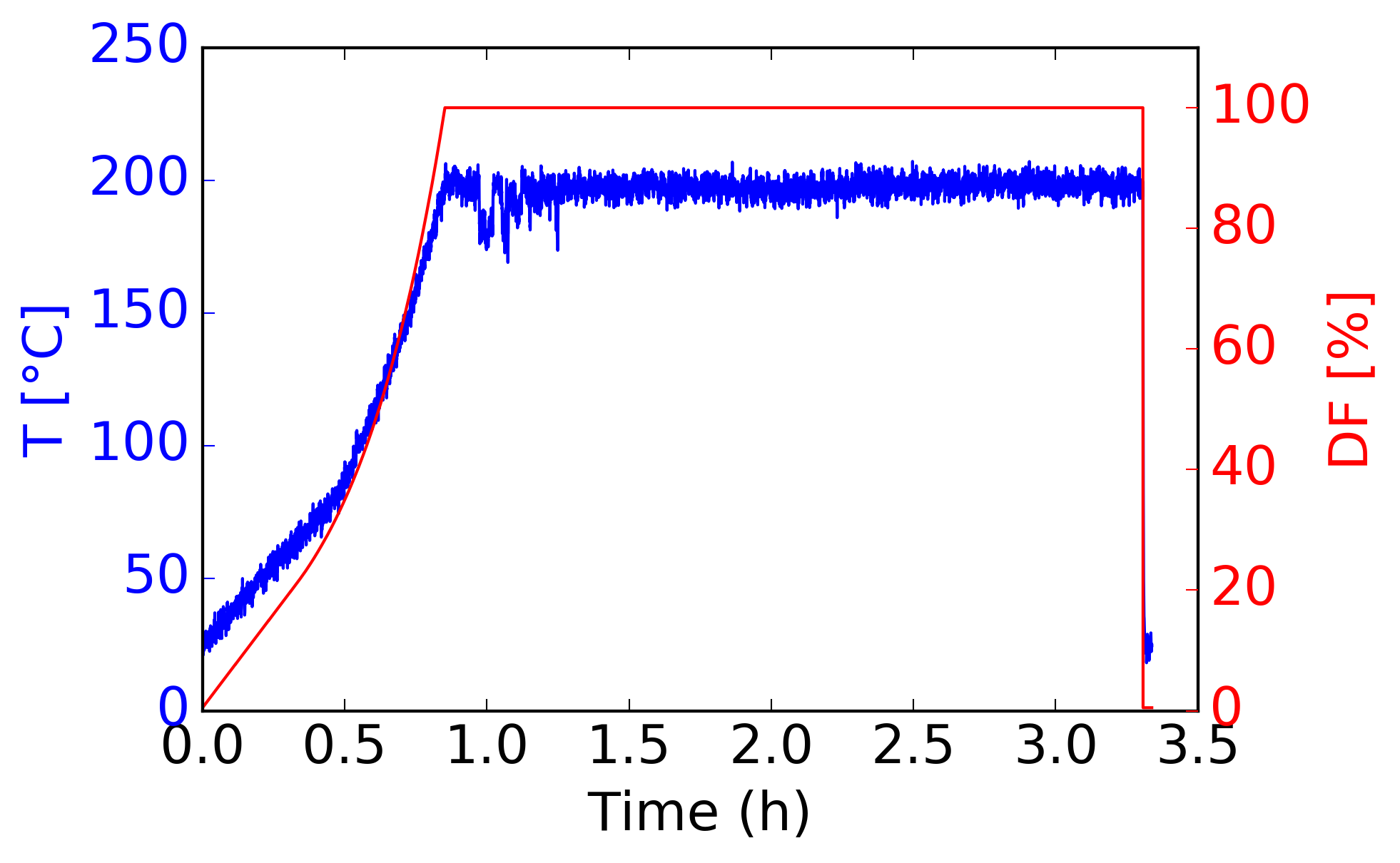}
	\caption{Temperature evolution of a Ta-Al sample while ramping up the electron beam Duty Factor (DF). This transitory ended with an abrupt beam interruption or "trip".}
	\label{fig:CTS_TaAl_Temperature}
\end{figure}

The tests carried out were intended to reproduce the operational scenario where a continuous beam is impinged at full power on the converter for 500 hours ($\approx$3 weeks). Moreover, 50 instantaneous beam interruptions were performed on the test samples to reproduce temperature cycles in future beam "trip" scenarios such as the one shown in Fig. \ref{fig:CTS_TaAl_Temperature}. The final acceptance decision is based on an initial visual inspection of the surface followed by scanning electron microscopy (SEM) and energy dispersive spectroscopy (EDX) on the material interface in sample cross sections. In order to save machine time, samples were visually inspected after the first 50 hours of irradiation to check the status of the irradiated surface. If its morphology changed, the "pre-acceptance" stage was considered failed and another sample would be irradiated.
\begin{figure*}
	%	\setbox1=\hbox{\includegraphics[height=3cm]{example-image-b}}
	\includegraphics[height=4.1cm]{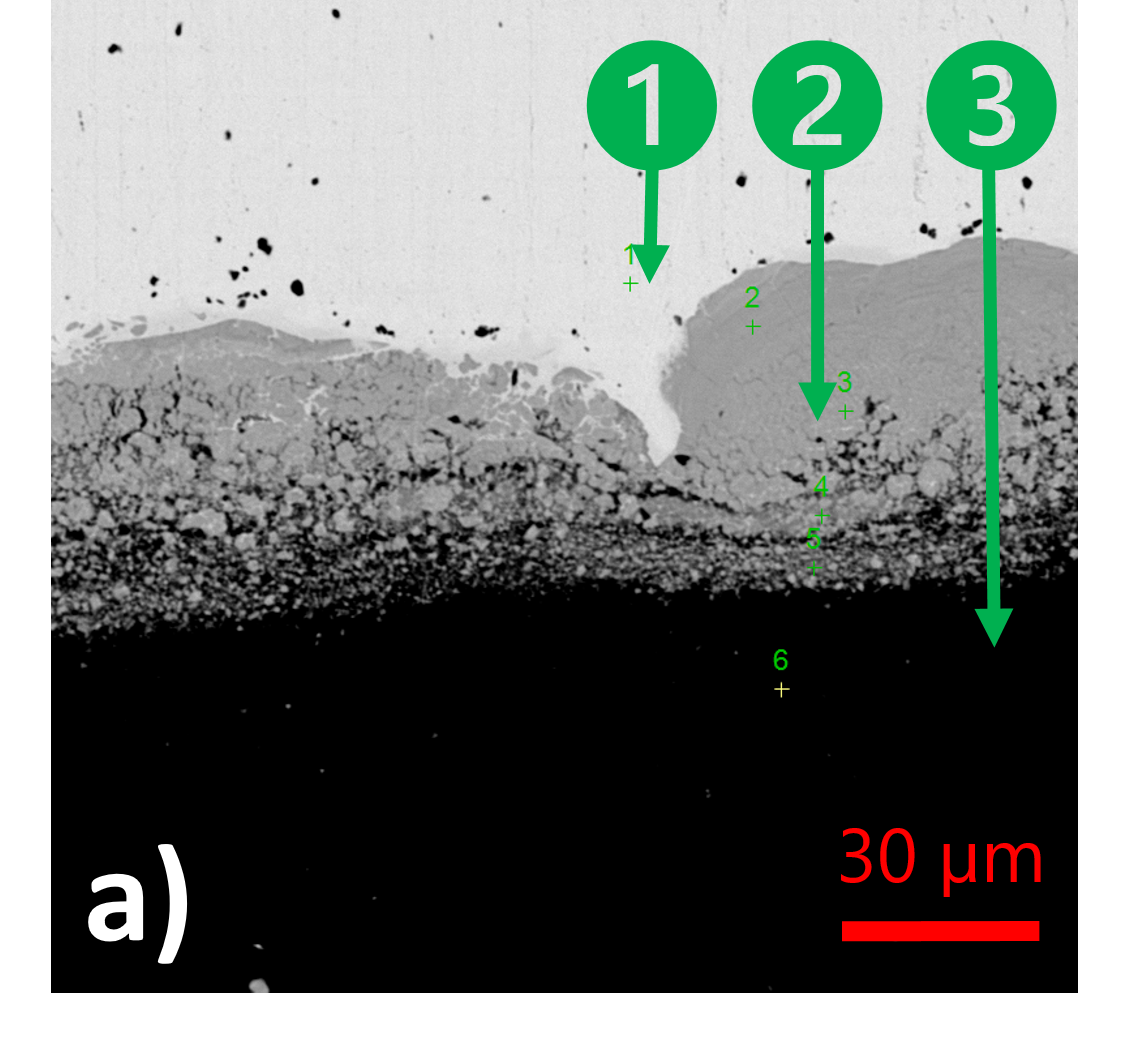}
	\includegraphics[height=4.1cm]{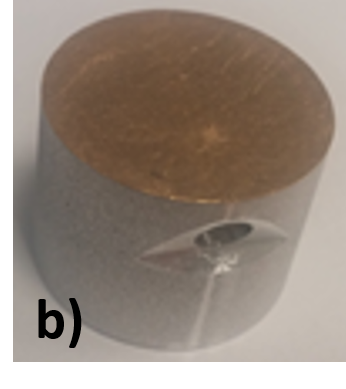}
	\includegraphics[height=4.1cm]{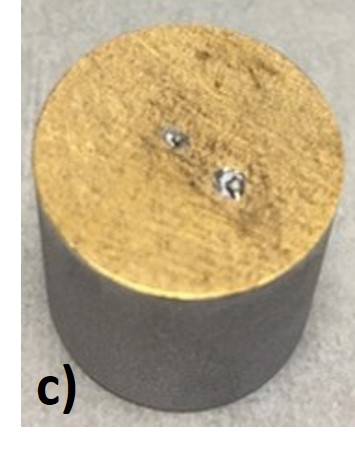}
	\includegraphics[height=4.1cm]{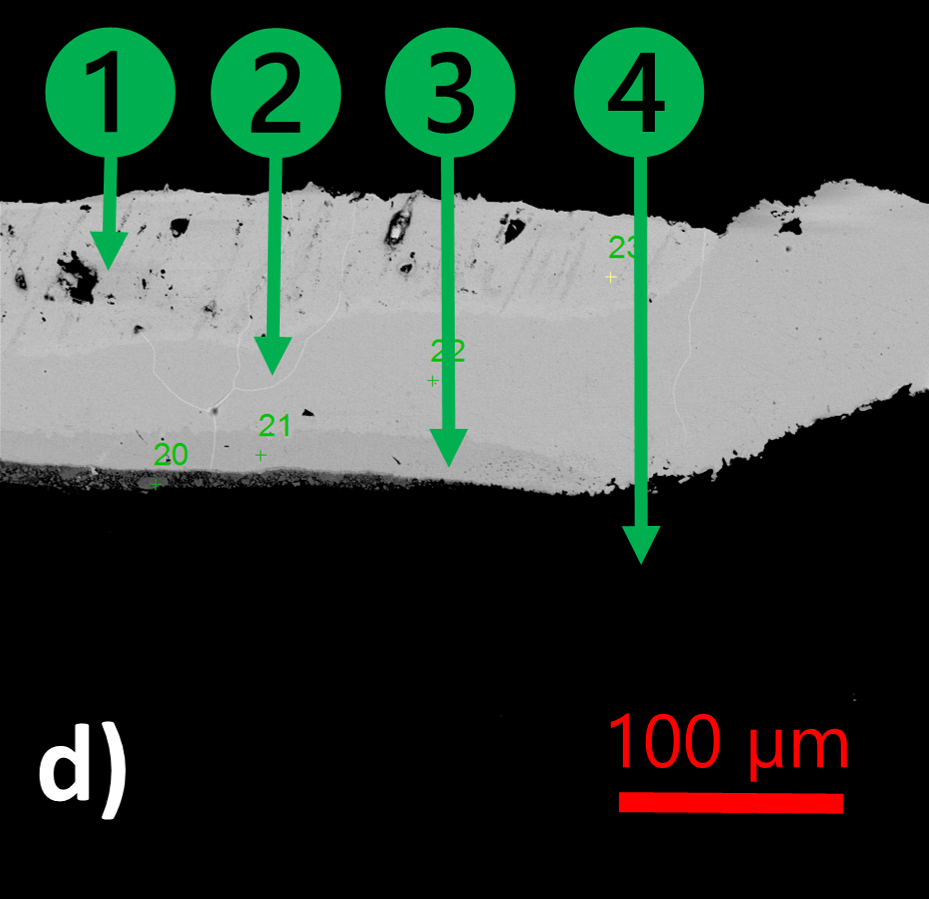}
	\includegraphics[width=\linewidth]{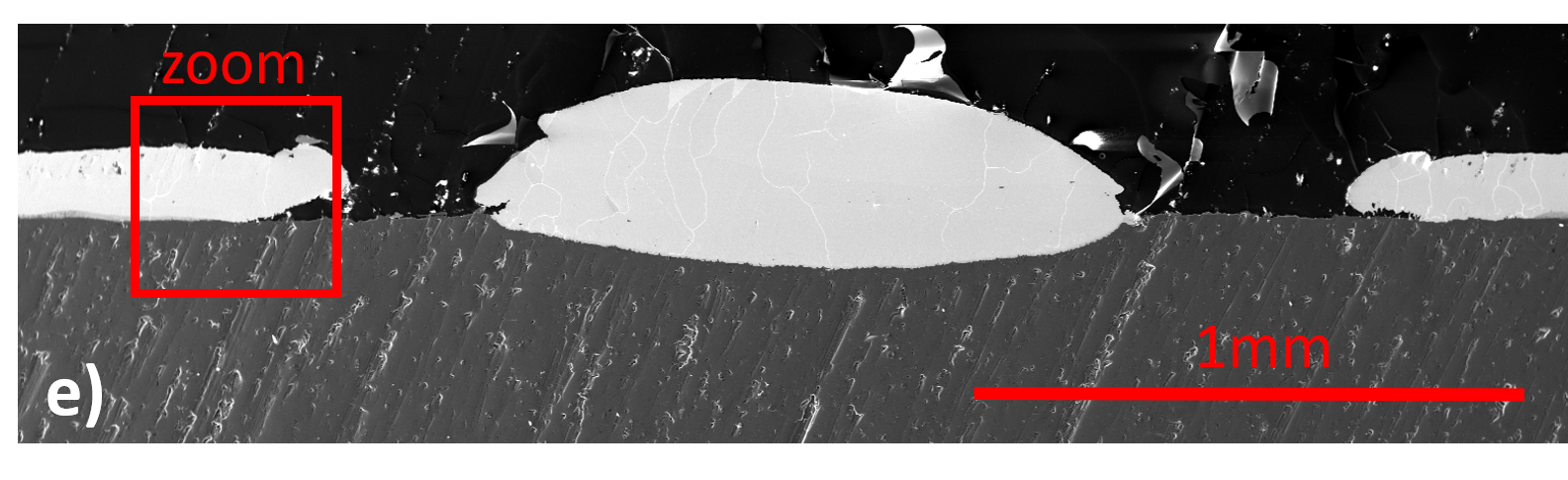}
	\caption{Sample 1 before (b) and after (c) irradiation for 50 hours. \\ Al \% in pristine sample (a): 1) 3.5\% 2) 51\% 3) 97\%  \\ Al \% in irradiated sample (d): 1) 1.5\% 2) 28\% 3) 32\% 4) 98\% \\ Zoomed-out SEM interface where gold gaps are evident after irradiation (e).}
	\label{fig:CTS_Samples_AuExpl}
\end{figure*}

\section{Test Results}
Three different gold-aluminum samples and two tantalum aluminum samples were prepared with different methods and the results are summarized in table~\ref{tab:test outcome}: 
\begin{itemize}
	\item Sample 1: Au-Al samples have been prepared by the company \textit{High Energy Metals Inc.} \cite{highenergymetals} by explosion bonding a 100 $\upmu$m gold layer onto a thick plate of aluminum-6061 alloy, and subsequently machined using the EDM technique into cylinders for testing.
	\item Sample 2: Au-Al samples have been prepared by the company \textit{SIFCOASC} \cite{sifcoasc} by selective brush plating a 500 $\upmu$m gold layer onto an aluminum-6061 alloy cylinder with a 3-4 $\upmu$m-thick nickel layer in between. The full layer was deposited during multiple steps of coating 70-100 $\mu m$ of gold and mechanical polishing. The nickel layer was inserted as a passivation layer which could potentially prevent migration of aluminum atoms into the gold layer and vice versa.
	\item Sample 3: Au-Al samples have been prepared at SLAC \cite{METSD} by electrodeposition of a  10 $\upmu$m gold layer onto an aluminum-6061 alloy cylinder with sub-$\upmu$m nickel and zinc layers in between. The nickel and zinc layers were inserted as potential passivation layers as mentioned for sample 2.
	\item Sample 4: Ta-Al samples have been prepared (by \cite{highenergymetals}) by explosion bonding a 1 mm thick tantalum layer onto a thick plate of aluminum-6061 alloy, and have been later machined into cylinders for testing.
	\item Sample 5: Ta-Al samples have been prepared (by \cite{highenergymetals}) by explosion bonding a 0.1 mm thick tantalum layer onto a thick plate of aluminum-6083 alloy, and subsequently machined into cylinders for testing.
\end{itemize}

\begin{figure*}
	%	\setbox1=\hbox{\includegraphics[height=3cm]{example-image-b}}
	\includegraphics[height=3.2cm]{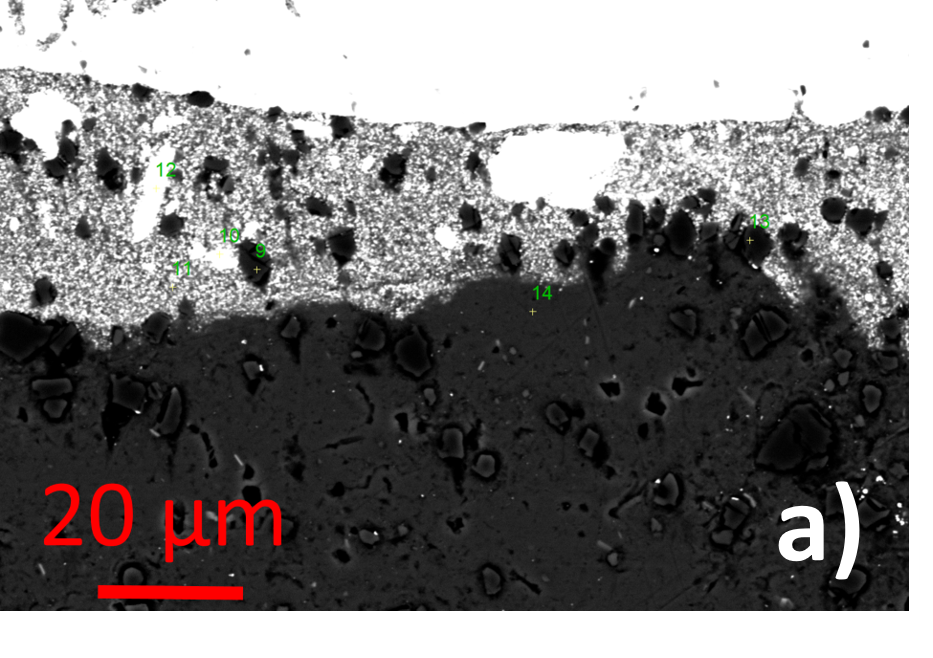}
	\includegraphics[height=3.2cm]{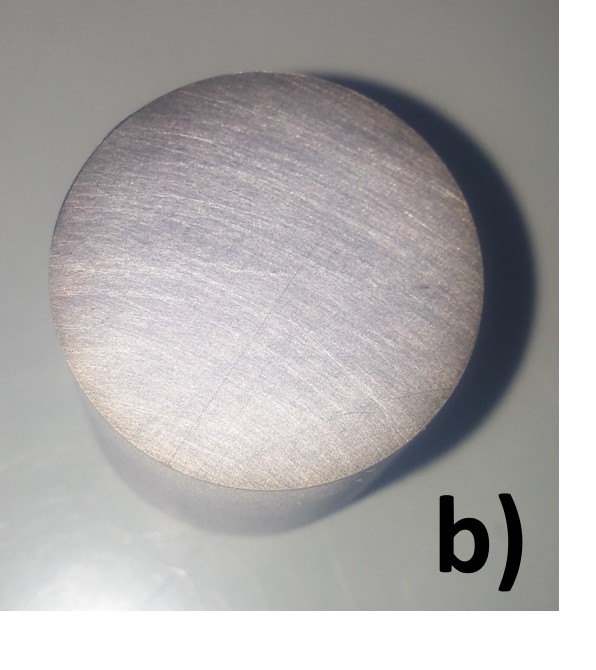}
	\includegraphics[height=3.2cm]{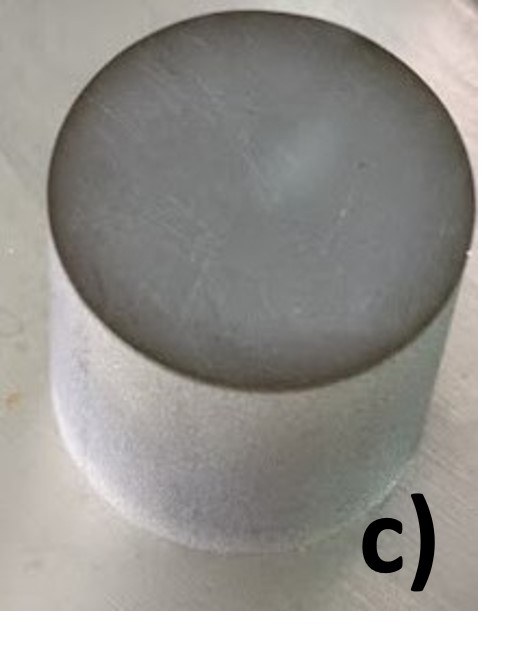}
	\includegraphics[height=3.2cm]{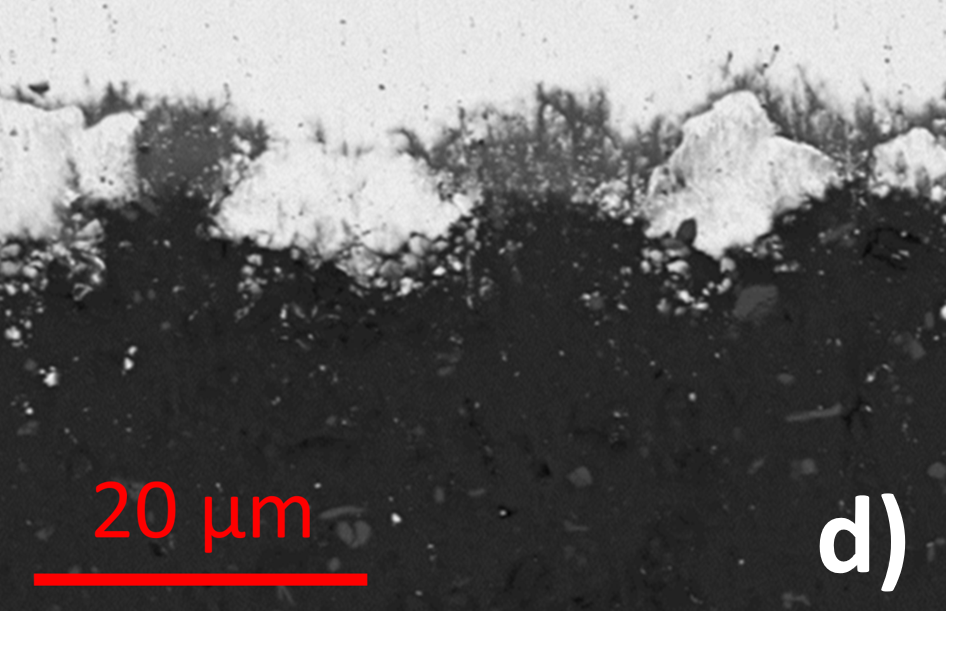}
	\centering
	\caption{Sample 4 interface before (a,b) and after (c,d) irradiation for 500 hours, with no major morphological or chemical change.}
	\label{fig:CTS_Samples_Ta}
\end{figure*}

\begin{figure*}
	\centering
	\includegraphics[width=\linewidth]{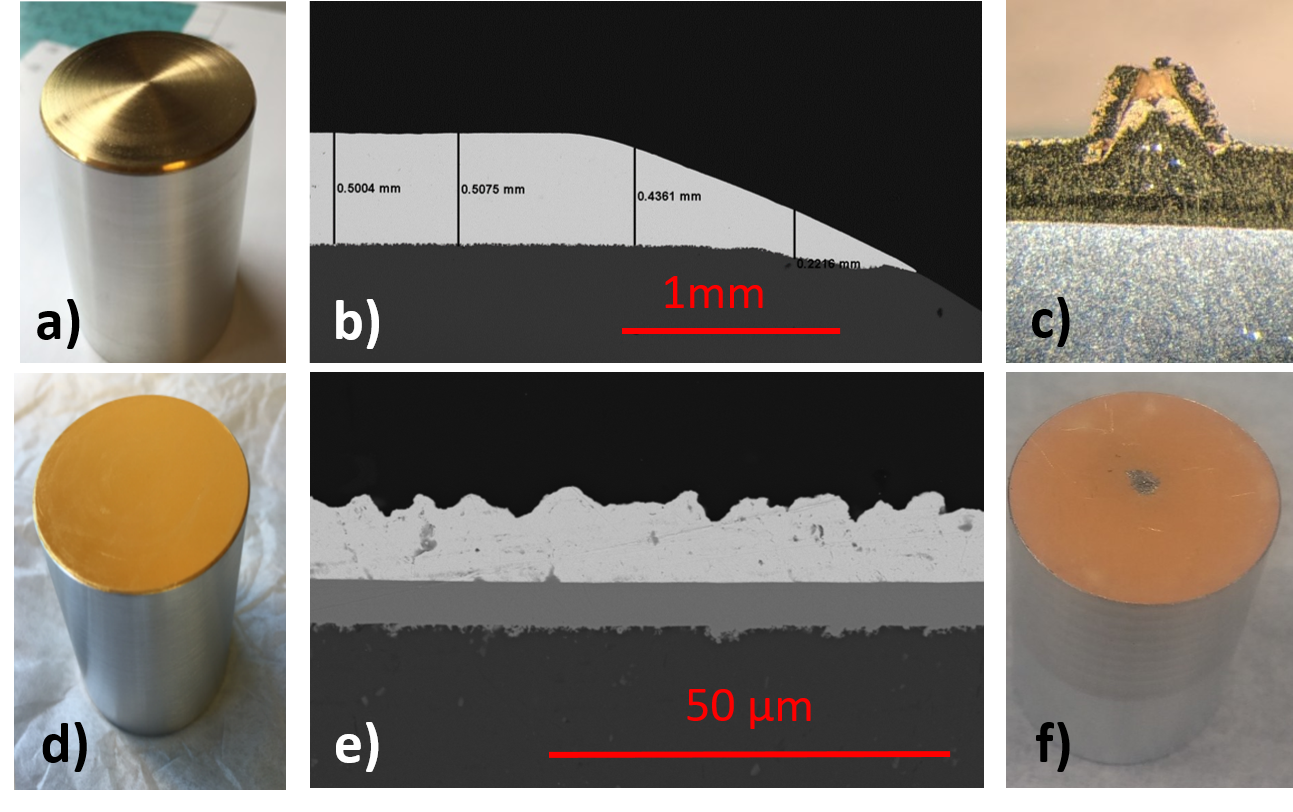}
	\caption{Sample 2 (top) and sample 3 (bottom). Pristine samples (a),(d) and their interface microstructure (b), (e). Pictures of both post-irradiation samples are shown in (c) and (f).}
	\label{fig:ElectroSamples}
\end{figure*}

\begin{table*}[t]
	\centering
	\caption{Samples irradiated and test outcomes. All samples from 1 to 5 were also exposed to 50 electron beam instant interruptions.}
	\label{tab:test outcome}
	\begin{tabular}{ c c c c c} 
		\toprule 
		Metal pairing 	& Bonding procedure & Irradiation time & Test outcome & High-Z layer thickness\\
			&	&	[h]&	&	 [$\upmu$m]\\
		\midrule
		1) Au-Al6061	& Explosion bonded & 50 & Fail & 100 \\[1ex]
		2) Au-Al6061	& Electroplated (Ni layer) & 50 & Fail & 500\\[1ex]
		3) Au-Al6061	& Electroplated (Ni + Zn layers) & 50 & Fail & 10 \\[1ex]
		4) Ta-Al6061	& Explosion bonded & 500 & Pass & 1000\\[1ex]
		5) Ta-Al6083	& Explosion bonded & 500 & Pass & 100\\[1ex]		
		\bottomrule
	\end{tabular}
\end{table*}

Fig. \ref{fig:CTS_Samples_AuExpl} shows the image of the explosion bonded gold sample 1 before and after the pre-acceptance test which consists of 50 hours of continuous irradiation plus 50 beam cycles each consisting of a slow (one hour long) duty factor ramp up followed by an instantaneous beam interruption (trip). The sample was cut and polished perpendicularly to the irradiated surface following irradiation to expose the Au-Al interface in correspondence with the damaged spot as shown in Fig. \ref{fig:CTS_Samples_AuExpl}e. SEM/EDX microscopy performed on sample~1 and the presence of a thick \ce{Au_{0.78}Al_{0.22}} phase - along with other phases as shown in Fig. \ref{fig:CTS_Samples_AuExpl}d  - is visible, in agreement with the Au-Al phase diagram \cite{Du1996} and the temperature and phase evolution kinetics described in \cite{Xu2011}. Similar damage is evident on the surface of the other two gold-aluminum samples (sample~2 and sample~3 in Fig. \ref{fig:ElectroSamples}c and \ref{fig:ElectroSamples}f). Moreover, temperature increase during irradiation was noticed on all the samples which showed morphological changes on their surface after 50 hours. Based on the similar temperature trends among all damaged samples it is concluded that the nickel and zinc interlayers do not prevent diffusion of gold and aluminum into each other. These damages are considered unacceptable for the survival of the converter, and the pre-acceptance test is considered failed for the three gold-aluminum samples. The evidence of resulting gaps in the top gold layer are detrimental for the lifetime of the online converter-target assembly and cannot be accepted.

Two tantalum-aluminum samples (samples 4 and 5) were irradiated for 500 hours at a power equivalent to 100~kW irradiation in the AETE design, and both have also been exposed to 50 beam cycles. Two different aluminum alloys were used in these tests, Al6061 and Al5086, since they were the two alloys considered as converter backing material due to their thermal and welding properties. Fig. \ref{fig:CTS_Samples_Ta} presents the SEM image on a pristine (a,b) and an irradiated (c,d) Ta-Al sample at the irradiation spot where no significant change in the interface is observed. These tests are considered passed, and the tantalum-aluminum metal pairing is selected as the AETE converter material. No further irradiations of the Au-Al pairing is foreseen, but further thermal tests will be conducted in an oven to investigate potential radiation enhanced phenomena.

\section{Conclusion}
The irradiations of several samples consisting of thin layers of high-Z material on aluminum alloy backings have been performed at the CTS using the 300 keV section of the ARIEL electron linac. Three gold-aluminum samples made with different production methods and protective interlayers have been irradiated for 50 hours and the morphology of the gold layer proved to be unstable at the chosen irradiation conditions. Two samples of explosion bonded tantalum on aluminum have been irradiated for 500 hours and no morphological or chemical changes in the interface have been observed by visual inspection or by SEM/EDX at the tantalum-aluminum interface. Tantalum-aluminum has been therefore selected as AETE converter material pairing.

\section*{Acknowledgements}
ARIEL is funded by the Canada Foundation for Innovation (CFI), the Provinces of AB, BC, MA, ON, QC, and TRIUMF. TRIUMF receives federal funding via a contribution agreement with the National Research Council of Canada. L.E. acknowledges support from the NSERC CREATE IsoSiM fellowship. The authors acknowledge Shane Koscielniak, Brandon Humphries and the rest of the ARIEL electron linac team for making these tests possible at the CTS, Brendan Cade for performing SEM/EDX of the sample interfaces and Lydia Young and SLAC for providing the gold sample 3. 



%% References

%\bibliographystyle{unsrt}

%\bibliographystyle{model1b-num-names}

%\section*{References}
\bibliographystyle{elsarticle-num}
\bibliography{../../../Literature/MendeleyFiles/library}

\begin{thebibliography}{10}
\expandafter\ifx\csname url\endcsname\relax
  \def\url#1{\texttt{#1}}\fi
\expandafter\ifx\csname urlprefix\endcsname\relax\def\urlprefix{URL }\fi
\expandafter\ifx\csname href\endcsname\relax
  \def\href#1#2{#2} \def\path#1{#1}\fi

\bibitem{ISACandARIEL}
J.~Dilling, R.~Kr{\"{u}}cken, L.~Merminga, {ISAC and ARIEL}, Springer,
  Dordrecht, Vancouver, 2014.
\newblock \href {http://dx.doi.org/https://doi.org/10.1007/978-94-007-7963-1}
  {\path{doi:https://doi.org/10.1007/978-94-007-7963-1}}.

\bibitem{Maloney2016}
J.~Maloney, R.~Baartman, M.~Marchetto,
  \href{https://www.sciencedirect.com/science/article/pii/S0168583X15012082?via{\%}3Dihub}{{New
  design studies for TRIUMF's ARIEL High Resolution Separator}}, Nuclear
  Instruments and Methods in Physics Research Section B: Beam Interactions with
  Materials and Atoms 376 (2016) 135--139.
\newblock \href {http://dx.doi.org/10.1016/J.NIMB.2015.11.023}
  {\path{doi:10.1016/J.NIMB.2015.11.023}}.
\newline\urlprefix\url{https://www.sciencedirect.com/science/article/pii/S0168583X15012082?via{\%}3Dihub}

\bibitem{Essabaa2003}
S.~Essabaa, J.~Arianer, P.~Ausset, O.~Bajeat, J.~Baronick, F.~Clapier,
  L.~Coacolo, C.~Donzaud, M.~Ducourtieux, S.~Gales, D.~Gardes, D.~Grialou,
  F.~Hosni, D.~Guillemaud-Mueller, F.~Ibrahim, T.~Junquera, C.~Lau, F.~L.
  Blanc, H.~Lefort, J.~L. Scornet, J.~Lesrel, A.~Mueller, J.~Obert, O.~Perru,
  J.~Potier, J.~Proust, F.~Pougheon, B.~Roussiere, N.~Rouviere, J.~Sauvage,
  O.~Sorlin, A.~Tkatchenko, D.~Verney, B.~Waast, L.~Rinolfi, G.~Rossat,
  D.~Forkel-Wirth, A.~Muller, G.~Bienvenu, J.-C. Bourdon, T.~Garvey,
  B.~Jacquemard, M.~Omeich, {Photo fission for the production of radioactive
  beams ALTO project}, Nuclear Instruments and Methods in Physics Research B
  204 (2003) 780--784.

\bibitem{Franchoo2015}
S.~Franchoo, \href{http://journals.jps.jp/doi/10.7566/JPSCP.6.020041}{{The Alto
  Tandem and Isol Facility at IPN Orsay}}, Proceedings of the Conference on
  Advances in Radioactive Isotope Science (ARIS2014) 020041 (2015) 1--6.
\newblock \href {http://dx.doi.org/10.7566/JPSCP.6.020041}
  {\path{doi:10.7566/JPSCP.6.020041}}.
\newline\urlprefix\url{http://journals.jps.jp/doi/10.7566/JPSCP.6.020041}

\bibitem{Cade2018}
B.~G. Cade, L.~Egoriti, A.~Gottberg, D.~Priessl,
  \href{http://stacks.iop.org/1742-6596/1067/i=3/a=032023?key=crossref.bf0c04a04fefaad2d539fb0493fe1af9}{{Thermal
  Design of a 100 kW Electron to Gamma Converter at TRIUMF}}, Journal of
  Physics: Conference Series 1067 (2018) 032023.
\newblock \href {http://dx.doi.org/10.1088/1742-6596/1067/3/032023}
  {\path{doi:10.1088/1742-6596/1067/3/032023}}.
\newline\urlprefix\url{http://stacks.iop.org/1742-6596/1067/i=3/a=032023?key=crossref.bf0c04a04fefaad2d539fb0493fe1af9}

\bibitem{Du1996}
Y.~Du, {Thermodynamic Modeling of the Al-Ta system}, Journal of Phase
  Equilibria 17~(4) (1996) 311--324.

\bibitem{Okamoto2005}
H.~Okamoto, {Al-Au (Aluminum-Gold)}, Journal of Phase Equilibria {\&} Diffusion
  26~(4) (2005) 391--393.
\newblock \href {http://dx.doi.org/10.1361/154770305X56999}
  {\path{doi:10.1361/154770305X56999}}.

\bibitem{Koscielniak2017}
S.~R. Koscielniak, Y.~M. Z. A. K. F. J. K. O. K. R.~L. {M. Laverty}, V.~Z. A.
  M. T. P. D. S. E. T. Z.~Y. {B. Waraich}, {TRIUMF ARIEL e-Linac Ready for 30
  MeV}, in: Proceedings, 8th International Particle Accelerator Conference
  (IPAC 2017): Copenhagen, Denmark, May 14-19, 2017, 2017, p. TUPAB022.
\newblock \href {http://dx.doi.org/10.18429/JACoW-IPAC2017-TUPAB022}
  {\path{doi:10.18429/JACoW-IPAC2017-TUPAB022}}.

\bibitem{highenergymetals}
\href{http://highenergymetals.com/}{{High Energy Metals, Inc}}.
\newline\urlprefix\url{http://highenergymetals.com/}

\bibitem{sifcoasc}
\href{http://www.sifcoasc.com/}{{SIFCO Applied Surface Concepts (ASC)}}.
\newline\urlprefix\url{http://www.sifcoasc.com/}

\bibitem{METSD}
{Mechanical Engineering {\&} Technical Services Division (METSD), SLAC National
  Accelerator Laboratory, Menlo Park, CA 94025}.

\bibitem{Xu2011}
H.~Xu, C.~Liu, V.~Silberschmidt, S.~Pramana, T.~White, Z.~Chen, V.~Acoff,
  \href{https://www.sciencedirect.com/science/article/pii/S0966979511002123}{{Intermetallic
  phase transformations in Au–Al wire bonds}}, Intermetallics 19~(12) (2011)
  1808--1816.
\newblock \href {http://dx.doi.org/10.1016/J.INTERMET.2011.07.003}
  {\path{doi:10.1016/J.INTERMET.2011.07.003}}.
\newline\urlprefix\url{https://www.sciencedirect.com/science/article/pii/S0966979511002123}

\end{thebibliography}

%% not want to use model1b-num-names.bst.





% \end{thebibliography}

\end{document}